\documentclass[12pt]{article}
\usepackage{amsmath}
\usepackage{amsthm}
\usepackage{amssymb}
\usepackage{latexsym}
\usepackage{subeqnarray}
\usepackage{graphicx}
\usepackage{bm}  
\newcommand{\nc}{\newcommand*}

\nc{\reff}[1]{(\ref{#1})}
\nc{\ds}{\displaystyle}
\nc{\ts}{\textstyle}
\nc{\nn}{\nonumber}

\usepackage[cp1251]{inputenc}
\inputencoding{cp1251}
\usepackage[T2B]{fontenc}
\usepackage[russian,english]{babel}
\begin{document}
\selectlanguage{english}

{\large\bf V.V. Borzov
\footnote{Department of Mathematics, St.Petersburg State University of Telecommunications, 22-1, Prospekt
Bolshevikov, St.Petersburg, 193232, Russia,\hfill\break  e-mail: borzov.vadim@yandex.ru}
E.V. Damaskinsky \footnote{Mathematical Department, VI(IT),  Zacharievskaya 22,
191123 Russia,\hfill\break e-mail: evd@pdmi.ras.ru}}
\bigskip

\begin{flushright} To the memory of our friend

V.D. Lyakhovsky\qquad\qquad\qquad
\end{flushright}
\bigskip

\begin{center}
{\Large\bf Realization by a differential operator
\\[0.2cm]
 of the annihilation operator for
\\[0.2cm]
generalized Chebyshev oscillator}
\end{center}
\bigskip

\begin{quote}
{\sc Abstract } We study a generalized Chebyshev oscillator \cite{1} associated with a point
interaction for the discrete Schr\"odinger equation.
Our goal is to find a realization of the annihilation operator
for this oscillator by a differential operator. This realization can
be used to obtain a differential equation for the corresponding
generalized Chebyshev polynomials \cite{2}.
This report is a continuation of our work  \cite{1}, \cite{3}.

\end{quote}

\vspace{1cm}


\section{Introduction}

This work continues the research of generalized Heisenberg algebras
\cite{4} connected with several orthogonal polynomial systems.
In \cite{3}, a general scheme was constructed for realization
of annihilation operator for these algebras by differential operator.
Note that with the exception of the standard case of Hermite polynomials, the differential
operator $\mathbf{A}$ appearing in such realizations has infinite order.

In the work \cite{3} one important special case of orthogonal polynomial systems, for which the
matrix of the  operator $A$ in $l^2(\mathrm{Z_{+}})$ has only off-diagonal
elements on the first upper diagonal different from zero, was considered.
The generalized Hermite polynomials \cite{5}, \cite{6} give us an example of
such an orthonormal  system.
In this work we consider another important case of orthogonal polynomial systems for which the
matrix of the operator A has only diagonal elements different from zero on all odd upper diagonals.
The generalized Chebyshev polynomials \cite{2}, \cite{7} give us an example of such an orthonormal  system.

\subsection{Generalized Chebyshev polynomials}

The generalized Chebyshev polynomials  $Ch_{n}(z;k;a), k\geq 1,$ are defined
by the recurrent relations:
\begin{multline}\label{01}
b_{n}Ch_{n+1}(z;k;a)+b_{n-1}Ch_{n-1}(z;k;a)=
zCh_{n}(z;k;a),\quad n\geq 0,\\
Ch_{0}(z;k;a)=1,\quad Ch_{-1}(z;k;a)=0,\qquad\qquad\qquad\qquad
\end{multline}
where $b_{k-1}=a,$ and $b_{n}=1,$ for $n\neq {k-1}.$
Using the expression obtained in \cite{3} for the polynomials connected
with relation \reff{01} (and associated Jacobi matrix), we have
\begin{equation*}
Ch_{n}(z;k;a)=
\!\!\!\sum_{m=0}^{Ent(\frac{n}{2})}\frac{(-1)^m}{\sqrt{[n]!}}b_{0}^{2m-n}\beta_{2m-1,n-1}z^{n-m},
\end{equation*}
where $\beta_{-1,n-1}=1,\, n\geq 0,$ and
\begin{equation*}
\beta_{2m-1,n-1}=\sum_{k_1=2m-1}^{n-1}\!\!\![k_1]!\sum_{k_2=2m-3}^{k_{1}-2}[k_2]!\cdots
\sum_{k_m=1}^{k_{m-1}-2}[k_m]!
\end{equation*}
for all $ m\geq 1.$ Here $[s]=\ds\frac{b_{s-1}^2}{b_{0}^2},$ and the integral part of $x$ denoted
by $Ent(x).$

As an example, we give the last formulas for the case $k=1$ (denote $\Psi_{n}(z)=Ch_{n}(z;1;a)$):
\begin{multline}\label{04}
\qquad\qquad\qquad\qquad\qquad\Psi_{0}(z)=1,\quad \Psi_{1}(z)=\frac{z}{a},\\
\Psi_{n}(z)=\frac{z^n}{a}-\frac{n+(a^{2}-2)}{a}z^{n-2}+\qquad\qquad\qquad\qquad\qquad\qquad\\
\sum_{m=2}^{Ent(\frac{n}{2})}(-1)^{m}\frac{(n-m-1)!(n+m(a^{2}-2))}{(n-2m)!m!a}z^{n-2m},\quad n\geq 2.
\end{multline}
In this paper, we will only consider  the case when $k=1$.

\subsection{Generalized Chebyshev oscillators (for the\\ case $\mathbf{k=1}$).}
Let $a>0$, $\mathcal{H}_a=L^2(\mathbb{R};\mu_{a})$ be a Hilbert space and $\lbrace\varphi_n(x)\rbrace_{n=0}^\infty$ be
a system of polynomials, which orthonormal with respect to the measure
$\mu_a,$ where
\begin{equation*}
d\mu_{a}(x)=\frac{1}{2\pi}
\left\{\begin{aligned} & \frac{a^2\sqrt{4-x^2}}{a^4-(a^2-1)x^2}dx,\quad\text{if}\quad |x|\leq 2,\\
&\quad  0,\quad\quad\quad\quad\quad\quad\quad\,\,\text{if}\quad |x|>2.\end{aligned}\right.
\end{equation*}
Then as follows from \cite{2} (see also \cite{7}), the polynomials $\varphi_n(x)$ are
generalized Chebyshev polynomials $\Psi_{n}(x)$ (for the case k=1) and the recurrent relations
\reff{01} takes the following form:
\begin{gather*}
a\Psi_{1}(x)=z\Psi_{0}(z),\quad \Psi_{2}(x)+a\Psi_{0}(x)=x\Psi_{1}(x),\nn\\
\Psi_{n+1}(x)+\Psi_{n-1}(x)=x\Psi_{n}(x),\quad n\geq 2,\\
\Psi_{0}(x)=1,\quad \Psi_{-1}(x)=0,\nn
\end{gather*}

In the work \cite{4} it was shown, that one can construct the oscillator-like algebra $\mathfrak{A}_{\psi}$
corresponding to this polynomial system.
The polynomials $\lbrace\Psi_n(x)\rbrace_{n=0}^\infty$ give the Fock basis
for this algebra $\mathfrak{A}_{\Psi}$ in the Fock space $\mathcal{H}_a$. The generators
$a_{\mu_a}^{+},a_{\mu_a}^{-},N_{\Psi}$ of the algebra
$\mathcal{A}_\Psi$ in this Fock representation acts as follows
\begin{equation}\label{06}
a_{\mu_a}^{+}\Psi_n=\sqrt{2}b_n\Psi_{n+1},\quad a_{\mu_a}^{-}\Psi_n=\sqrt{2}b_{n-1}\Psi_{n-1},
\quad N_{\Psi}\Psi_n=n\Psi_n,
\end{equation}
where
\begin{equation*}
b_{-1}=0,\quad b_0=a,\quad b_n=1,\quad  n\geq1.
\end{equation*}
Let $I$ be the identity  operator in the Hilbert space
$\mathcal{H}_a.$
We define $B_{\Psi}(N_{\Psi})$ as an operator-valued
function defined by the relations
\begin{equation*}
B_{\Psi}(N_{\Psi})\Psi_{n}=b_{n-1}^{2}\Psi_{n},\quad B_{\Psi}(N_{\Psi}+I)\Psi_{n}=b_{n}^{2}\Psi_{n},
\quad n\geq 0.
\end{equation*}
Then the generalized Chebyshev oscillator algebra $\mathfrak{A}_{\Psi}$ is generated by operators
$a_{\Psi}^{\pm},$  $N_{\Psi}$ and $I$ satisfying the relations
\begin{gather*}
a_{\mu_a}^{-}a_{\mu_a}^{+}\Psi_{n}=2B_{\Psi}(N_{\Psi}+I),\quad
a_{\mu_a}^{+}a_{\mu_a}^{-}\Psi_{n}=2B_{\Psi}(N_{\Psi}),\quad\\
[N_{\Psi},a_{\mu_a}^{\pm}]=\pm a_{\mu_a}^{\pm},
\end{gather*}
and by the commutators of these operators.

\subsection{Statement of a problem }
The main purpose of the present work is to find the coefficients $a_{ls},$
under which the operator $\mathbf{A}$ defined by the relation $a_{\mu_a}^{-}=\sqrt{2 }\mathbf{A}$ takes the form
\begin{equation}\label{07}
\mathbf{A}=\sum_{s=1}^{\infty}{\sum_{l=0}^{s-1}a_{ls}z^l \frac{d^s}{dz^s}}.
\end{equation}
Denote by $A=\{a_{ls}\}$\quad $(1\leq s<\infty,\quad 0\leq l\leq (s-1))$  the matrix of the operator $\mathbf{A}$.
Using the definition of the operator $\mathbf{A}$ and (\ref{06}) we
obtained the relations
\begin{equation}\label{08}
\mathbf{A}\Psi_n=b_{n-1}\Psi_{n-1},\quad n\geq 1,\quad \mathbf{A}\Psi_0=0.
\end{equation}
Then, substituting the relations \reff{04} and \reff{07} in \reff{08}, we get the main
relation for to find elements $a_{ls}$ of the matrix $A$:
\begin{multline}\label{09a}
\sum_{s=1}^{n}{\sum_{l=0}^{s-1}}a_{ls}( \frac{n!}{(n-s)!}z^{n+l-s}-(n-2+a^2)\frac{(n-2)!}{(n-2-s)!}z^{n+l-s-2}+\\
\sum_{m=2}^{Ent(\frac{n}{2})}(-1)^{m}
\frac{(n-m-1)!(n+m(a^{2}-2))}{(n-2m-s)!m!}z^{n+l-2m-s})=\\
z^{n-1}- (n-3+a^2)z^{n-3}+\\
\sum_{m=2}^{Ent(\frac{n}{2})}(-1)^{m}
\frac{(n-m-2)!(n-1+m(a^{2}-2))}{(n-2m-1)!m!}z^{n-1-2m}.
\end{multline}
By equating the coefficients for $z^{n-2k}$ in the left and right parts of the identity (\ref{09a}),
we obtain the following equations for finding the coefficients $a_{s-2t,s}$, under the condition
$s\geq 2t\geq 2$:
\begin{gather}\label{10}
\sum_{m=1}^{k}{(-1)^{m-1}\frac{n+(m-1)(a^2-2)}{(m-1)!}}\!\!\!
\sum_{s=2(k-m+1)}^{n-2(m-1)}\!\!\!\!\!\!\!\!\!{a_{s-2(k-m+1),s}\frac{(n-m)!}{(n-s-2(m-1))!}}\!=\!0.
\end{gather}
Similarly, using the coefficients for degrees $z^{n-2k-1}$, we obtain the equations for determining the
coefficients $a_{s-2t-1,s}$, under the condition $s\geq1,\, t\geq 0.$
\begin{multline}\label{11}
\sum_{m=1}^{k+1}\!(-1)^{m-1}\frac{n+(m-1)(a^2-2)}{(m-1)!}\times \\
\sum_{s=2(k-m+1)+1}^{\,\,n-2(m-1)}{a_{s-2(k-m+1)-1,s}\frac{(n-m)!}{(n-s-2(m-1))!}} \\
=(-1)^k \frac{(n-k-2)!(n-1-2k+ka^2)}{(n-1-2k)!k!}.
\end{multline}

\section{Calculating elements of the matrix $A$ for the annihilation operator}
\subsection{Calculating elements of even overdiagonals of the matrix $A$}
In this paragraph we will prove that the elements
$a_{n-2k,n}$  of even overdiagonals of the annihilation  operator matrix  $\mathbf{A}$  equal to zero:
\begin{equation}\label{12}
a_{n-2k,n}=0  ,\quad  k\geq 1,\quad n\geq 2k.
\end{equation}
First, we show that
\begin{equation}\label{13}
a_{0,2q}=0,\quad  q\geq 1.
\end{equation}
To do this, we use equality  (\ref{10}) for $n=2p,\, k=p :$
\begin{gather*}
a_{0,2p}\frac{2p!}{0!}-a_{0,2p-2}(2p-2+a^2)\frac{(2p-2)!}{1!}+\\
a_{0,2p-4}(2p-4+2a^2)\frac{(2p-4)!}{2!}+...+(-1)^{p-1}a_{0,2}a^2=0.
\end{gather*}
Consistently using the last equality for all $ p=1,2, ...,q ,$  we get (\ref{13}).

Next, we will prove the formula
\begin{equation}\label{14}
a_{1,1+2q}=0,\quad  q\geq 1.
\end{equation}
We use equality  (\ref{10})
 for $n=2p+1 , k=p :$
\begin{gather*}
a_{0,2p}\frac{(2p+1)!}{1!}+a_{1,2p+1}\frac{(2p+1)!}{0!}-\\
(2p-1+a^2)(a_{0,2p-2}\frac{(2p-1)!}{1!1!}+a_{1,2p-1}\frac{(2p-1)!}{0!1!})+\\
(2p-3+2a^2)(a_{0,2p-4}\frac{(2p-2)!}{1!2!}+a_{1,2p-3}\frac{(2p-2)!}{0!2!})+\\
...+(-1)^{p-1}(1+pa^2)(a_{0,2}\frac{p!}{1!(p-1)!}+a_{1,3}\frac{p!}{0!(p-1)!})=0.
\end{gather*}
Consistently using the last equality for all $ p=1,2, ...,q ,$  we get
(\ref{14}).

Finally, we consider the general case
\begin{equation}\label{15}
a_{t,t+2q}=0,\quad  q\geq 1,\quad  t\geq 0.
\end{equation}
It is obvious that $a_{t,t+2q}$ passes to $a_{n-2k,n}$  at $q=k,\, t=n-2k.$
 It is easy to see from formula (\ref{10}) that all coefficients
$a_{t,t+2q}$ are expressed only in terms of coefficients $a_{s,s+2p}$
under conditions $0\leq s < t,\, p\leq q$ and $s=t,\, p<q.$ Then, using
the proven formulas (\ref{13}),\, (\ref{15}) and gradually increasing the
indices $p$ and $s$ is not difficult to prove the validity of equality (\ref{12}).

\subsection{Calculating elements of odd overdiagonals of the matrix  $A$}
The main result of this work is the following formula:
\begin{equation}\label{16}
a_{l,l+2k+1}=\frac{(-1)^{l+1}}{(l+2k+1)!}((1-\delta_{l,0})C_k C_{l+2k}^{2k+1}-
C_{l+2k+1}^{2k+1}P_{k,2k+2}(a)),l,k\geq 0,
\end{equation}
where $C_k=\ds\frac{C_{2k}^k}{k+1}$ are Catalan numbers,  $\delta_{0,0}=1$ and $\delta_{l,0}=0$ for
 $l>0.$ The polynomials $P_{k,2k+2}(a)$ are defined as follows
\begin{multline}\label{17}
\qquad\qquad\qquad P_{0,2}(a)=a^2;\quad P_{1,4}(a)=a^4;\quad P_{2,6}(a)=a^6+a^4; \\
P_{k,2k+2}(a)=a^{2k+2}+\sum_{i=1}^{k-1}{\beta_{k,i}}a^{2(k-i+1)}\quad k\geq 3;\\
\beta_{k,1}=k-1,\, k\geq 2;\quad
\beta_{k,i}=\frac{(k+1)(k+2)...(k+i-1)(k-i)}{i!},\quad \\
2\leq i\leq k-2;\quad \beta_{k,k-1}=\beta_{k,k-2}=C_{k-1},\quad k\geq 4. \qquad\qquad
\end{multline}

From (\ref{11})we have
\begin{multline*}
\qquad\qquad a_{n-(2k+1),n}=\frac{1}{n!}\left(-\sum_{s=2k+1}^{n-1}{a_{s-(2k+1),s}\frac{n!}{(n-s)!}}+\right.\\
\sum_{m=2}^{k+1}(-1)^{m-2}\!\!\!\!\!\!\!\!\!\sum_{s=(2k+1)-2(m-1)}^{n-2(m-1)}\!\!\!\!\!\!\!\!\!\!\!\!
a_{s-(2k+1-2(m-1)),s}\frac{(n-m)!( n+(m-1)(a^2-2))}{(n-s-2(m-1))!(m-1)!}+\\
\left.(-1)^k\frac{(n-k-2)!(n-1-2k+ka^2)}{(n-1-2k)!k!}\right).\qquad\qquad
\end{multline*}
Replacing the  indices $l=n-(2k+1),t=s-(2k+1)+2(m-1),$ we get
\begin{multline}\label{19}
a_{l,l+(2k+1)}=\frac{1}{(l+2k+1)!}\left(-\sum_{t=0}^{l-1}{a_{t,t+(2k+1)}\frac{(l+2k+1)!}{(l-t)!}}+\right.\\
\sum_{m=2}^{k+1}(-1)^{m-2}\sum_{t=0}^{l}
a_{t,t+(2k+1-2(m-1))}\times \\
\frac{(l+2k+1-m)!( l+2k+1+(m-1)(a^2-2))}{(l-t)!(m-1)!}+\\
\left.(-1)^k\frac{(l+k-1)!(l+ka^2)}{l!k!}\right).
\end{multline}

In the remaining part of this article, we will prove the formula (\ref{16}) by
induction, using the relation (\ref{19}). As the base of induction, we take the
case $l=0,k=0$. Obviously, the formula (\ref{16}) is valid, since
\begin{equation}\label{20}
a_{0,0+1} = a^2=P_{0,2}.
\end{equation}
 We note that  the coefficients $a_{l,l+(2k+1)}$  are expressed
using the formula (\ref{19})  by elements $a_{u,v}$ with "smaller" numbers,
i.e. for $u\leq l, v\leq l+(2k+1)$  and $u+v<2l+2k+1$.
Then the induction transition consists in the fact that assuming the correctness of
formula (\ref{16}) for all coefficients $a_{u,v}$  standing on the right side of
equality (\ref{19}), we must prove the fulfillment of formula (\ref{16}) also
for the left side of equality (\ref{19}).
To prove this statement, it is sufficient to check the equality of all coefficients for the same
powers of $a^2$ polynomials in both parts of the relation (\ref{19}), which arise when
substituting in (\ref{19}) expressions for all $a_{t,t+(2s+1)}$ by the formula (\ref{16}).

As can be seen from the definition (\ref{07}) of operator $\mathbf{A}$, the matrix $A$
of this operator is an upper triangular matrix. It is convenient to start the proof of
formula (\ref{16}) by considering the "boundary" non-zero elements of the matrix $A$,
that is, elements $a_{l,l+1}$ ($k=0,l>0$) standing on the first overdiagonal
of the matrix $A$  and elements $a_{0,2k+1}$ ($k\geq0,l=0$) standing on the first
row of the matrix $A$ .

\section{Two special cases of the formula (\ref{16})}
\subsection{The case $k=0,\, l>0$}
From (\ref{19}) ($k=0,\, l>0$) we get
\begin{equation}\label{21}
(l+1)!\,a_{l,l+1}=-\sum_{t=0}^{l-1}{a_{t,t+1}\frac{(l+1)!}{(l-t)!}}+
(-1)^0\frac{(l-1)!l}{l!0!}.
\end{equation}
Note that the formula (\ref{16}) for $k=0,\, l>0$ has the form
\begin{equation}\label{22}
a_{l,l+1}=\frac{(-1)^{l+1}}{(l+1)!}\left(C_l^1-C_{l+1}^1 P_{0,2}(a)\right).
\end{equation}
Substituting (\ref{22}) in (\ref{21}) and using (\ref{20}) we get
\begin{multline*}
(-1)^{l+1}\left(\frac{l(l+1)!}{(l+1)!}-a^2\frac{(l+1)!}{l!}\right)= \\
-\sum_{t=0}^{l-1}(-1)^{t+1}tC_{l+1}^{t+1}
+a^2\sum_{t=0}^{l-1}{(-1)^{t+1}(t+1)C_{l+1}^{t+1}}+1.
\end{multline*}
Moving the first and second summands from the right part to the left part of the last equality, we get
\begin{equation}\label{23}
\sum_{t=0}^l(-1)^{t+1}tC_{l+1}^{t+1}-a^2\sum_{t=0}^l{(-1)^{t+1}(t+1)C_{l+1}^{t+1}}=1.
\end{equation}
For proof (\ref{23}), it is sufficient to check the fulfillment of the following two equalities
\begin{equation}
\sum_{t=0}^l(-1)^{t+1}tC_{l+1}^{t+1}=1,\quad\text{и}\quad\sum_{t=0}^l{(-1)^{t+1}(t+1)C_{l+1}^{t+1}}=0.
\end{equation}
The first of these equalities follows from the identity (\ref{49}) proved below for $q=0$.
To prove the second equality, write the left part of it as a sum
\begin{equation*}
\sum_{t=0}^l(-1)^{t+1}tC_{l+1}^{t+1}+\sum_{t=0}^l(-1)^{t+1}C_{l+1}^{t+1}=1-1=0.
\end{equation*}
Thus, the formula (\ref{16}) for the case $k=0,\, l>0$ is proved.

\subsection{The case $l=0,k\geq0$ (proof of the formula (\ref{17}))}
First of all, consider the special case $l=0,\, k\geq 0.$ Note that for $l=0$ ,
the first term in brackets in the right part of equality (\ref{16}) disappear,
i.e. equality (\ref{16}) takes the form
\begin{equation} \label{26}
a_{0,2k+1}=\frac{1}{(2k+1)!}
C_{2k+1}^{2k+1}P_{k,2k+2}(a), \quad k\geq 0.
\end{equation}
The formula is valid for $k=0$, since as already mentioned (see (\ref{20})) $a_{0,1}=a^2$,
if we assume that \[P_{0,2}(a)=a^2.\] Next, we can rewrite formula (\ref{19}) for the case $l=0,k\geq 1$ (
Note that for $l=0$ , the first term in brackets in the right part of equality (\ref{19})  disappear)
\begin{multline}\label{27}
a_{0,2k+1}=\frac{1}{(2k+1)!}\left(\sum_{m=2}^{k+1}[(-1)^{m-2}
a_{0,2k+1-2(m-1)}\times \right.\\
\left.\frac{(2k+1-m)!( 2k+1+(m-1)(a^2-2))}{0!(m-1)!}]+
(-1)^k\frac{(k-1)!ka^2}{0!k!}\right).
\end{multline}
To check formula  (\ref{27}), it is sufficient to check the equality of all coefficients
with the same powers of $a^2$ of the polynomials that arise when substituting in both parts of the relation
(\ref{27}) expressions for all $a_{0,2k+1-2(m-1)}$ by formula (\ref{26}).
In other words, we need to check the following identity.
\begin{multline}\label{28}
P_{k,2k+2}(a)=\sum_{m=2}^{k+1}\left[\rule[-10pt]{0pt}{26pt}(-1)^{m-2}P_{k-m+1,2(k-m+1)+2}(a)\times \right. \\
\left.\frac{(2k+1-m)!(2k+1+(m-1)(a^2-2))}{(2(k-m+1)+1)!(m-1)!}\right]+(-1)^ka^2,
\quad k\geq 1.
\end{multline}

Consider the cases $k=1$ and $k=2$. When $k=1$ we have
\begin{equation}\label{29}
P_{1,4}(a)=(-1)^2 P_{0,2}\frac{1!(3+a^2-2)}{1!}-a^2=a^2(1+a^2)-a^2=a^4.
\end{equation}
When $k=2$ we have
\begin{equation}\label{30}
P_{2,6}(a)=P_{1,4}(a)\frac{3!(5+a^2-2)}{3!1!}-P_{0,2}(a)\frac{2!(5+2a^2-4)}{1!2!}+a^2=a^6+a^4.
\end{equation}
Note that, using (\ref{29}) and (\ref{30}), it is not difficult to obtain from (\ref{28})
the following polynomial $P_{3,8}(a)$
\begin{equation}\label{31}
P_{3,8}(a)=a^8+2(a^6+a^4).
\end{equation}
Finally, for $k\geq 2$, we will look for an expression for the polynomial $P_{k,2k+2}(a)$ in the following form
\begin{equation}\label{32}
P_{k,2k+2}(a)=a^{2k+2}+\sum_{i=1}^{k-1}\beta_{k,i}a^{2(k-i+1)}.
\end{equation}
Comparing the expressions (\ref{30}) and (\ref{31}) for the polynomials $P_{2,6}(a)$
and $P_{3,8}(a)$ with the corresponding expressions obtained from (\ref{32}) for $k=2$
and $k=3$, respectively, we have
\begin{equation}\label{33}
\beta_{2,1}=C_1=1,\quad \beta_{3,1}=\beta_{3,2}=C_2=2.
\end{equation}

Next we will use (\ref{32}) for $k\geq3$.
To check formula (\ref{28}) for $k\geq 3$, we need to rewrite it in the following form
\begin{multline}\label{34}
P_{k,2k+2}(a)=\sum_{m=2}^{k-2}[(-1)^{m-2}P_{k-m+1,2(k-m+1)+2}(a)\times\\
\frac{(2k+1-m)!(2k+1+(m-1)(a^2-2))}{(2(k-m+1)+1)!(m-1)!}]+\\
+(-1)^{k-3}P_{2,6}(a)\frac{(2k+1-k+1)!(2k+1+(k-2)(a^2-2))}{(4+1)!(k-2)!}+\\
+(-1)^{k-2}P_{1,4}(a)\frac{(k+1)!(2k+1+(k-1)(a^2-2))}{(2+1)!(k-1)!}+\\
+(-1)^{k-1}P_{0,2}(a)\frac{k!(2k+1+k(a^2-2))}{1!k!}+(-1)^ka^2,\quad k\geq 1.
\end{multline}
Substituting in (\ref{34}) instead of the  polynomials $P_{p,2p+2}(a)$
their expressions by the formula (\ref{32}) for $p\geq 3$, we have
\begin{multline}\label{35}
a^{2k+2}+\sum_{i=1}^{k-1}\beta_{k,i}a^{2(k-i+1)}=\sum_{m=2}^{k-2}[(-1)^{m-2}(a^{2(k-m+1)+2}+ \\
+\sum_{j=1}^{k-m+1-3}\beta_{k-m+1,j}a^{2(k-m+1-j+1)}+
\beta_{k-m+1,k-m+1-2}(a^6+a^4))\times\\
\frac{(2k+1-m)!(2k+1+(m-1)(a^2-2))}{(2(k-m+1)+1)!(m-1)!}]+\\
(-1)^{k-3}P_{2,6}(a)\frac{(2k+1-k+1)!(2k+1+(k-2)(a^2-2))}{(4+1)!(k-2)!}+\\
(-1)^{k-2}P_{1,4}(a)\frac{(k+1)!(2k+1+(k-1)(a^2-2))}{(2+1)!(k-1)!}+\\
(-1)^{k-1}P_{0,2}(a)\frac{k!(2k+1+k(a^2-2))}{1!k!}+(-1)^ka^2,\quad k\geq 1.
\end{multline}
By equating the coefficients for the same degrees
of $a^2$ in both parts of the relation (\ref{35}) we get the validity of formulas
 (\ref{17}) for the values $\beta_{k,i}$.

Let's start by finding $\beta_{k,k-2}$  (coefficient for $a^6$) and
$\beta_{k,k-1}$ (coefficient for $a^4$):
\begin{multline}\label{67}
\qquad\qquad\beta_{k,k-2}=(-1)^{k-2}C_{k+1}^{k-2} +
\sum_{t=1}^{k-2}(-1)^{k-2-t}C_t C_{k+2+t}^{k-1-t}=C_{k-1},\,\,\\
\beta_{k,k-1}=(-1)^{k-1}C_{k}^{k-1} +
\sum_{t=0}^{k-2}(-1)^{k-2-t}C_t C_{k+1+t}^{k-1-t}=C_{k-1}.\qquad
\end{multline}
For the proof of formulas (\ref{67}), see Appendix 1.
Thus, we proved that the following equalities are valid for $k\geq2$
\begin{equation}\label{36}
\beta_{k,k-2}=\beta_{k,k-1}=C_{k-1}.
\end{equation}
By equating the coefficients for $a^{2(k-i+1)}$ for $k\geq 3$ and $1\leq i\leq k-2$,
we get the following system of equations for the quantities $\beta_{p,q}$
\begin{multline}\label{37}
\beta_{k,i}=(-1)^{i-1}C_{2k-i}^i + (-1)^iC_{2k-i-1}^i+(-1)^{i-1}
\beta_{k-i,1}C_{2k-i}^{i-1}+\\
\sum_{m=2}^{i}(-1)^{m-2}\left(\beta_{k-m+1,i-m+1}C_{2k-m+1}^{m-1}+
\beta_{k-m+1,i-m+2}C_{2k-m+1}^{m-2}\right).
\end{multline}
Let's start by calculating the value of $\beta_{k,1}$. We will prove the following formula
\begin{equation}\label{38}
\beta_{k,1}=k-1.
\end{equation}
by induction.
The above equation $\beta_{3,1}=2$ serves as the base of induction. From the formula (\ref{37}) we have
\begin{equation*}
\beta_{k,1}=(-1)^0 C_{2k-1}^1+(-1)^1 C_{2k-2}^1 +(-1)^0 \beta_{k-1,1}C_{2k-1}^0= 1+\beta_{k-1,1}.
\end{equation*}
The last equality provides an inductive transition: if $\beta_{k-1,1}=k-2$,
then $\beta_{k,1}=1+k-2=k-1$. Thus the formula (\ref{38}) is proved.

Consider also the case $\beta_{k,2}$. From (\ref{37})  for $k\geq 3$ and $i=2$  we have
\begin{gather*}
\beta_{k,2}=-C_{2k-2}^2 + (-1)^2 C_{2k-3}^2+(-1)^{0}
(\beta_{k-1,1}C_{2k-1}^{1}+\beta_{k-1,2}C_{2k-1}^{0})+\\
(-1)^1\beta_{k-2,1}C_{2k-2}^{1}=\beta_{k-1,2}+k-1.
\end{gather*}
The last equality provides an inductive transition and above equation $\beta_{3,2}=2$
serves as the base of induction. Then
\begin{equation}\label{40}
\beta_{k,2}=(k-1)+(k-2)+... +2= \frac{(k+1)(k-2)}{2!},\quad k\geq 3.
\end{equation}
This coincides with the formula (\ref{17}) for $i=2$.

Taking into account (\ref{20}), (\ref{29}), (\ref{30}), (\ref{32}),
(\ref{36}) and (\ref{38}), to complete the proof of formulas (\ref{17}) it remains to
check that for coefficients
$\beta_{k,i}$ defined by the equality (\ref{37}) the following relation is true.
\begin{multline}\label{41}
\qquad\beta_{k,i}= C_{k+i-1}^{k-1}-C_{k+i-1}^{k}=\frac{k-i}{i}C_{k+i-1}^{k-1}=\\
\frac{(k+1)(k+2)...(k+i-1)(k-i)}{i!}. \qquad\qquad
\end{multline}
We will prove (\ref{41}) by induction. In this case, (\ref{33}) and (\ref{40})
are considered as the base of induction. The inductive transition means that the coefficients
$\beta_{k,i}$ calculated by the formula (\ref{37}) must satisfy the equality (\ref{41}),
provided that all the coefficients $\beta_{p,q}$ included in the right part (\ref{37})
satisfy the equality (\ref{41}).

So, to prove (\ref{41}) it is sufficient to check the validity of the following equality
\begin{multline}\label{42}
(-1)^iC_{2k-i-1}^i+(-1)^{i-1}C_{2k-i}^i + (-1)^{i-1}
C_{k-i-1}^1 C_{2k-i}^{i-1}+\\
\sum_{m=2}^{i}(-1)^{m-2}\left[(C_{k+i-2m+1}^{k-m}-C_{k+i-2m+1}^{k-m+1})C_{2k-m+1}^{m-1}+\right.\\
\left.(C_{k+i-2m+2}^{k-m}-C_{k+i-2m+2}^{k-m+1})C_{2k-m+1}^{m-2})\right]=C_{k+i-1}^{k-1}-C_{k+i-1}^{k},
\quad 2\leq i\leq k-2.
\end{multline}
Denote by $S(m;k,i)$ the general term of the sum in \reff{42}:
\begin{multline*}
S(m;k,i)=(-1)^{m-2}\left[(C_{k+i-2m+1}^{k-m}-C_{k+i-2m+1}^{k-m+1})C_{2k-m+1}^{m-1}+\right.\\
\left.(C_{k+i-2m+2}^{k-m}-C_{k+i-2m+2}^{k-m+1})C_{2k-m+1}^{m-2})\right].
\end{multline*}
It is easy to see that
\begin{eqnarray*}
S(1;k,i)&=&-C_{k+i-1}^{k-1}+C_{k+i-1}^{k},\\
S(i+1;k,i)&=&(-1)^{i-1}(C_{2k-i}^i + C_{k-i-1}^1 C_{2k-i}^{i-1}),\\
S(i+2;k,i)&=&(-1)^iC_{2k-i-1}^i.
\end{eqnarray*}
Then the relation (\ref{41}) can be rewritten as
\begin{equation}\label{43}
\sum_{m=1}^{i+2}S(m;k,i)=0.
\end{equation}
The validity of this equality was tested in the program of symbolic computation "Mathematica".
Thus, the the relation (\ref{41}), is proved and therefore the validity of the
formula (\ref{17}) is established.

\section{Checking the equality of first-degree polynomials}

Now we consider the case $l>0,k>0$. By equating polynomials of the first degree from $a^2,$
arising when substituting in (\ref{19}) expressions for all
$a_{t,t+(2s+1)}$ by the formula (\ref{16}), we get
\begin{multline}\label{44}
\frac{(-1)^{l+1}}{(l+2k+1)!}C_k C_{l+2k}^{2k+1}
=\frac{1}{(l+2k+1)!}\times\\
\left\{\sum_{t=0}^{l-1}(\delta_{t,0}-1)\frac{(-1)^{t+1}}{(t+2k+1)!}C_k
C_{t+2k}^{2k+1}\frac{(l+2k+1)!}{(l-t)!}+\right.\\
\sum_{m=2}^{k+1}(-1)^{m-2}\sum_{t=0}^{l}(1-\delta_{t,0})
[\frac{(-1)^{t+1}}{(t+2(k-m+1)+1)!}C_{k-m+1} C_{t+2(k-m+1)}^{2(k-m+1)+1}\times\\
\frac{(l+2k+1-m)!( l+2k+1+(m-1)(a^2-2))}{(l-t)!(m-1)!}]+\\
\left. (-1)^k\frac{(l+k-1)!(l+ka^2)}{l!k!}\right\}.
\end{multline}
So, we need to proof the validity of this equality. To do this, just check the following two relations:
\begin{multline}\label{45}
\sum_{m=2}^{k+1}(-1)^{m-2}\frac{(l+2k+1-m)!(m-1)}{(m-1)!}\sum_{t=1}^{l}C_{k-m+1}C_{t+2(k-m+1)}^{2(k-m+1)+1}\times\\
\frac{(-1)^{t+1}}{(l-t)!(t+2(k-m+1)+1)!}=(-1)^{k+1} \frac{(l+k-1)!k}{l!k!},
\end{multline}
and also
\begin{multline}\label{46}
(-1)^{l+1}(1-\delta_{l,0})C_k C_{l+2k}^{2k+1}=
\sum_{t=0}^{l-1}(\delta_{t,0}-1)
\frac{(-1)^{t+1}}{(t+2k+1)!}C_k
C_{t+2k}^{2k+1}\frac{(l+2k+1)!}{(l-t)!}+\\
\sum_{m=2}^{k+1}(-1)^{m-2}\sum_{t=0}^{l}(1-\delta_{t,0})
\left[\frac{(-1)^{t+1}}{(t+2(k-m+1)+1)!}C_{k-m+1} C_{t+2(k-m+1)}^{2(k-m+1)+1}\times \right.\\
\left.\frac{(l+2k+1-m)!( l+2k+1+(m-1)(-2))}{(l-t)!(m-1)!}\right]+
(-1)^k\frac{(l+k-1)!l}{l!k!}.
\end{multline}
First, we will check the equality \reff{45}, which means that in this case the
coefficient for $a^2$ on the right side of (\ref{16}) is zero.

We rewrite equality (\ref{45}) with replacing the  index of summation \hfill\break
$m=k+1-q$ ($m>1$)
\begin{multline}\label{47}
\sum_{q=0}^{k-1}(-1)^{k-q-1}\frac{(l+k+q)!}{(k-q-1)!}
\sum_{t=1}^{l}C_{q}C_{t+2q}^{2q+1}\frac{(-1)^{t+1}}{(l-t)!(t+2q+1)!}=\\
(-1)^{k-1} C_{l+k-1}^{k-1}.
\end{multline}

Now we will compare equality (\ref{47}) for $s=l+k$ with the identity (which was proved in Appendix 2):
\begin{equation}\label{48}
\sum_{q=0}^{k-1}(-1)^{k-q-1}C_q C_{s+q}^{k-1-q}=(-1)^{k-1} C_{s-1}^{k-1},
\quad 1\leq k \leq s.
\end{equation}
It is obvious that to prove equality (\ref{45}), it is sufficient to check that the coefficients
for all $C_q$ in the left parts of relations (\ref{47}) and (\ref{48}) coincide, that is
\begin{equation}\label{49}
\sum_{t=1}^{l}(-1)^{t+1}C_{t+2q}^{1+2q}C_{l+2q+1}^{t+2q+1}=1.
\end{equation}
Let's replace the variable $\tau=t+2q+1$

\begin{equation}\label{50}
\sum_{\tau=2q+2}^{l+2q+1}(-1)^{\tau-2q}C_{\tau-1}^{1+2q}C_{l+2q+1}^{\tau}=1.
\end{equation}
Since \[C_{\tau-1}^{1+2q}C_{l+2q+1}^{\tau}= C_{l+2q+1}^{l} \frac{l}{\tau}
C_{l-1}^{\tau-2q-2}\] (\ref{50}) can be rewritten as
\begin{equation*}
C_{l+2q+1}^{l}l \sum_{\tau=2q+2}^{l+2q+1}(-1)^{\tau-2q}\frac{C_{l-1}^{\tau-2q-2}}{\tau}=1.
\end{equation*}
By changing the summation index again $s=\tau-2(q+1)$ we get
\begin{equation*}
C_{l+2q+1}^{l}l \sum_{s=0}^{l-1}(-1)^s \frac{C_{l-1}^s}{s+2(q+1)}=1.
\end{equation*}
Using the formula (4.2.2.45) from \cite{8}, we have
\begin{equation*}
C_{l+2q+1}^{l}l\frac{(l-1)!}
{\prod_{s=0}^{l-1}{(s+2q+2)}}=\frac{(l+2q+1)!}{(l+2q+1)!}=1,
\end{equation*}
so equality (\ref{50}) is true. This proves the validity of the relationship (\ref{45}).

Let's go to the equality check (\ref{46}). For convenience, we will rewrite equality
(\ref{46}) in the following form.
\begin{equation}\label{52}
A=B_1+B_2,
\end{equation}
where
\begin{equation*}
A=(-1)^{l+1}(1-\delta_{l,0})C_k C_{l+2k}^{2k+1}+
\sum_{t=0}^{l-1}(1-\delta_{t,0})
\frac{(-1)^{t+1}}{(t+2k+1)!}C_k
C_{t+2k}^{2k+1}\frac{(l+2k+1)!}{(l-t)!},
\end{equation*}
and
\begin{multline}\label{54}
B_1=\sum_{m=2}^{k+1}(-1)^{m-2}\sum_{t=0}^{l}(1-\delta_{t,0})
\left[\frac{(-1)^{t+1}}{(t+2(k-m+1)+1)!}\times \right. \\
\left.C_{k-m+1} C_{t+2(k-m+1)}^{2(k-m+1)+1}
\frac{(l+2k+1-m)!( l+2k+1)}{(l-t)!(m-1)!}\right];\\[7pt]
B_2=\sum_{m=2}^{k+1}(-1)^{m-2}\sum_{t=0}^{l}(1-\delta_{t,0})
\left[\frac{(-1)^{t+1}}{(t+2(k-m+1)+1)!}\times\right.\\
\left. C_{k-m+1} C_{t+2(k-m+1)}^{2(k-m+1)+1}
\frac{(l+2k+1-m)!(m-1)(-2)}{(l-t)!(m-1)!}\right]+(-1)^k\frac{(l+k-1)!l}{l!k!}.
\end{multline}
Let's start by calculating the value of A. It is obvious that the first term can be included in the second sum, i.e.
\begin{multline}\label{55}
A=\sum_{t=0}^{l}(1-\delta_{t,0})(-1)^{t+1}C_k C_{t+2k}^{2k+1}C_{l+2k+1}^{t+2k+1}=\\
C_{l+2k+1}^{2k+1}C_k\sum_{t=0}^{l}(1-\delta_{t,0})(-1)^{t+1}C_l^t\frac{t}{t+2k+1}=\\
C_{l+2k+1}^{2k+1}C_k\left[\sum_{t=1}^{l}{(-1)^{t+1}C_l^t}-(2k+1)\sum_{t=1}^{l}{\frac{(-1)^{t+1}}{t+2k+1}C_l^t}\right].
\end{multline}
From the properties of binomial coefficients, it follows that the first term in square brackets is equal to 1, and the second,
after replacing $t=\tau+1$ takes the form
\begin{equation*}
-l\left(\sum_{\tau=0}^{l-1}\frac{(-1)^\tau}{\tau+1}C_{l-1}^\tau-
\sum_{\tau=0}^{l-1}\frac{(-1)^\tau}{\tau+2(k+1)}C_{l-1}^\tau\right).
\end{equation*}
Using the formula (4.2.2.45) from \cite{8}, we  can continue the equality (for the second term in square brackets)
\[-l\left(\frac{(l-1)!}{l!}-\frac{(l-1)!}{(2k+2)(2k+3)...(2k+2+l-1)}\right).\]
Substituting the resulting expression instead of the second term in the right
side of the expression (\ref{55}) for the value A, we get
\begin{equation}\label{56}
A=-C_k C_{l+2k+1}^{2k+1}+C_k C_{l+2k+1}^{2k+1}+C_kl!\frac{(2k+l+1)!}{(2k+1)!(2k+2)...(2k+l+1)l!}=C_k.
\end{equation}

Now let's start calculating $B_1$. By replacing the index $m=k+1-q$ in the right
part of the expression for $B_1$ (see (\ref{54})), in the same way as in (\ref{45})
when getting (\ref{47}), we get the following expression for $B_1$.
\begin{equation}\label{57}
B_1=(l+2k+1)\sum^{q=k-1}_{0}(-1)^{k-q-1}\frac{(l+k+q)!}{(k-q)!}
\sum_{t=1}^{l}C_{q}C_{t+2q}^{2q+1}\frac{(-1)^{t+1}}{(t+2q+1)!(l-t)!}.
\end{equation}
The coefficient for $C_q$ in this sum has the form
\begin{equation*}
(-1)^{k-q-1}C_{l+k+q}^{k-q-1}\frac{l+2k+1}{k-q}\sum_{t=1}^{l}(-1)^{t+1}
C_{t+2q}^{1+2q}C_{l+2q+1}^{t+2q+1}.
\end{equation*}
Note that according to (\ref{49}) the last sum is equal to 1.
Substituting the found coefficient values for $C_q$ in the formula (\ref{57}), we find (denote $s=l+k$)
\begin{multline*}
B_1=\sum_{q=0}^{k-1}(-1)^{k-q-1}C_qC_{s+q}^{k-q-1}\frac{s+k+1}{k-q}=\\
\sum_{q=0}^{k-1}(-1)^{k-q-1}C_qC_{s+q}^{k-q-1}+
 \sum_{q=0}^{k-1}(-1)^{k-q-1}C_qC_{s+q+1}^{k-q}=\\
 \sum_{q=0}^{k-1}(-1)^{k-q-1}C_qC_{s+q}^{k-q-1}-
\sum_{q=0}^{k}(-1)^{k-q}C_qC_{s+q+1}^{k-q}+ C_k.
\end{multline*}
Using the identity (\ref{48})(and its analog) for the first and second sums,
we get the final expression for the value $B_1$
\begin{equation}\label{59}
B_1=(-1)^{k-1}C_(s-1)^{k-1}+(-1)^{k-1}C_s^k + C_k.
\end{equation}

Finally, we calculate the value $B_2$. From the comparison of formulas (\ref{54})
and (\ref{47}), it is obvious that the first sum in the right part (\ref{54})
coincides with the left part (\ref{47}) multiplied by the multiplier (-2).
Then from (\ref{54}) and (\ref{47}) follows the equality.
\begin{multline*}
B_2= (-1)^{k-1}C_{s-1}^{k-1}(-2)+(-1)^k C_{s-1}^k=\\
=(-1)^k(2C_{s-1}^{k-1}+C_{s-1}^k)=(-1)^k C_{s}^k+ (-1)^k C_{s-1}^{k-1}.
\end{multline*}

Let's check the validity of equality (\ref{52}), using (\ref{59}),
(\ref{56}) and (\ref{47}). We have
\begin{multline*}
A=C_k,\quad B_1+B_2=(-1)^{k-1}C_{s-1}^{k-1}+(-1)^{k-1}C_s^{k} + C_k +\\ (-1)^{k} C_{s}^{k}+ (-1)^{k} C_{s-1}^{k-1}=C_k.
\end{multline*}
Hence the equity of equality (\ref{52}), and hence of equality (\ref{46})  and (\ref{44}) are proved.

\section{Checking the equality of polynomials of degree higher than the first}
Assuming $l>0$ and equating polynomials of degree higher than the first from $a^2,$
arising when substituting in (\ref{19}) expressions for all
$a_{t,t+(2s+1)}$ by the formula (\ref{16}), we get
\begin{multline}\label{61}
\frac{(-1)^{l+1}}{(2k+1)!l!}P_{k,2k+2}(a)=\\
\frac{1}{(l+2k+1)!}
\left\{-\sum_{t=0}^{l-1}(\frac{(-1)^{t+1}}{(2k+1)!t!}P_{k,2k+2}(a)
\frac{(l+2k+1)!}{(l-t)!}+\right.\\
\sum_{m=2}^{k+1}(-1)^{m-2}\sum_{t=0}^{l}
\left[\frac{(-1)^{t+1}}{(2(k-m+1)+1)!t!}P_{k-m+1,2(k-m+1)+2}(a)\right.\\
\left.\left.\frac{(l+2k+1-m)!( l+2k+1+(m-1)(a^2-2))}{(l-t)!(m-1)!}\right]\right\}.
\end{multline}
So, we need to check the validity of this equality.
For convenience, we will rewrite equality (\ref{61}) in the following form.
\begin{equation}\label{62}
\tilde{A}=\tilde{B},
\end{equation}
where
\begin{equation*}
\tilde{A}=\frac{(-1)^{l+1}}{(2k+1)!l!}P_{k,2k+2}(a)+
\sum_{t=0}^{l-1}\frac{(-1)^{t+1}}{(2k+1)!t!}\frac{P_{k,2k+2}(a)}{(l-t)!},
\end{equation*}
and
\begin{multline*}
\tilde{B}=\sum_{m=2}^{k+1}(-1)^{m-2}\sum_{t=0}^{l}
\left[\frac{(-1)^{t+1}}{(2(k-m+1)+1)!t!}P_{k-m+1,2(k-m+1)+2}(a)\times \right.\nn\\
\left.\frac{(l+2k+1-m)!( l+2k+1+(m-1)(a^2-2))}{(l-t)!(m-1)!}\right].
\end{multline*}
Let's start by calculating the value of $\tilde{A}$. It is obvious that the first term can be included in the second sum, i.e.
\begin{equation*}
\tilde{A}=\sum_{t=0}^{l}\frac{(-1)^{t+1}}{(2k+1)!t!}\frac{P_{k,2k+2}(a)}{(l-t)!}
=\frac{P_{k,2k+2}(a)}{l!(2k+1)!}\sum_{t=0}^{l}(-1)^{t+1}C_l^t=0.
\end{equation*}
Finally, we calculate the value $\tilde{B}$. To do this, calculate the coefficient for the polynomial $P_{k-m+1,2(k-m+1)+2}(a)$:
\begin{equation*}
(-1)^{m-2}\frac{(l+2k+1-m)!( l+2k+1+(m-1)(a^2-2))}{l!(2(k-m+1)+1)!(m-1)!}\sum_{t=0}^{l}(-1)^{t+1}C_l^t=0.
\end{equation*}
Then the equality (\ref{62}) is obvious, and hence the identity (\ref{61}),  is true.
This proves the validity of formula (\ref{19}) for all $k>0$ and $l>0$. Therefore,
formula (\ref{16}) is fully proved.

\section{Concluding remarks}
Recently in the theory of orthogonal polynomials
linear differential operators of arbitrary order are used.
Let us give some examples.

Firstly, there are systems of orthogonal polynomials, which only satisfy
a linear differential equation of infinite order \cite{10}.

Secondly, it is well known \cite{11} that any linear transformation
$$T : \mathbb{C}[x] \rightarrow \mathbb{C}[x]$$
has a differential operator  representation
\begin{equation*}
T= \sum_{n=0}^{\infty}\frac{Q_n(x)}{n!}D^n,
\end{equation*}
where $D$ denoted differentiation so that $D^nf(x)=f^{(n)}(x)$ and where
the $Q_n(x)$ are complex polynomials.
Such transformations that preserve or shrink the location
of the complex zeros of polynomials
is a recent object of study, motivated by the Riemann Hypothesis.

Further examples, as well as an extensive bibliography, are available in the monograph \cite{12}.
Thus the importance of linear differential operators
of infinite order
in the study of orthogonal polynomials needs no further emphasis.

In this paper, we obtain an realization of the annihilation operator
$a_{\mu_a}^{-}=\sqrt{2 }\mathbf{A}$ for the oscillator-like system, associated with a system of generalized Chebyshev polynomials
$Ch_{n}(z;1;a)$, by a differential operator of infinite order.
This operator has the form
\begin{equation*}
\mathbf{A}=\sum_{s=1}^{\infty}{\sum_{l=0}^{s-1}a_{ls}z^l \frac{d^s}{dz^s}}.
\end{equation*}
Formulas for calculating the coefficients $a_{ls}$ are obtained.
To illustrate, let's write out the beginning  of first few rows and columns of an infinite coefficient matrix
\begin{equation*}
\setcounter{MaxMatrixCols}{10}
A\!=\!\left\lceil
\begin{matrix}
0&a^2&0&\frac{a^4}{3!}&0&\frac{a^6+a^4}{5!}&0&\frac{a^8+2(a^6+a^4)}{7!}&\cdots\\
0&0&\frac{1-2a^2}{2!}&0&\frac{1-4a^4}{4!}&0&\frac{2-6(a^6+a^4)}{6!}&0&\cdots\\
0&0&0&-\frac{2-3a^2}{3!}&0&-\frac{4-10a^4}{5!}&0&-\frac{12-21(a^6+a^4)}{7!}&\cdots\\
0&0&0&0&\frac{3-4a^2}{4!}&0&\frac{10-20a^4}{6!}&0&\cdots\\
0&0&0&0&0&-\frac{4-5a^2}{5!}&0&-\frac{20-35a^4}{7!}&\cdots\\
0&0&0&0&0&0&\frac{5-6a^2}{6!}&0&\cdots\\
0&0&0&0&0&0&0&-\frac{6-7a^2}{7!}&\cdots\\
\hdotsfor{9}\\
\hdotsfor{9}\\
&&&&&&&&
\end{matrix}
\right\rceil
\end{equation*}

We hope that similar representations of ladder operators can
be useful in the study of generalized Heisenberg algebras,
related to systems of orthogonal polynomials. For example, when
obtaining differential equations for the corresponding polynomials
by the method proposed in the work of the authors \cite{3}.

\bigskip

{\bf Acknowledgements} Authors are grateful to I. K. Litskevich for assistance in performing some calculations.
\bigskip

\section*{Appendix 1\footnote{When proving this and the following formulas we will use a variant
"Snake Oil" of the generating function method \cite{9}.}}
{$\mathbf{1}$} Let's prove the first of the relations \reff{67}, namely
\begin{equation}\label{B01}
(-1)^{k-2}C_{k+1}^{k-2}+\sum_{t=1}^{k-2} (-1)^{k-2-t}C_t C_{k+2+t}^{k-1-t}=C_{k-1}.
\end{equation}
We transform this equality by entering the right hand side under the sign of the sum,
and take into account that for $t=k-1$ we have $(-1)^{k-2-t}=(-1)$
and $C_{k+2+t}^{k-1-t}=C_{2k+1}^{0}=1$. As a result, \reff{B01} goes to
\begin{equation*}
(-1)^{k-2}C_{k+1}^{k-2}+\sum_{t=0}^{k-1} (-1)^{k-2-t}C_t C_{k+2+t}^{k-1-t}=0
\end{equation*}
or, after shortening by $(-1)^{k-2}$ to
\begin{equation}\label{B02}
C_{k+1}^{k-2}+\sum_{m=0}^{k-2} (-1)^{m+1}C_{m+1} C_{k+m+3}^{k-m-2}=0,
\end{equation}
where $m=t-1$.  Since for $m\geq k-1$ we have $C_{k+m+3}^{k-m-2}=0$, then after shortening by $(-1)$,
we can rewrite \reff{B02} in the form
\begin{equation}\label{B03}
\sum_{m=0}^{\infty}(-1)^m C_{m+1}C_{k+m+3}^{2m+5}=C_{k+1}^{3}.
\end{equation}
Denoting $n=k+3$, we get
\begin{equation}\label{B04}
\sum_{m=0}^{\infty}(-1)^m C_{m+1}C_{n+m}^{2m+5}=C_{n-2}^{n-5}.
\end{equation}
This is the relation we will prove. To calculate the sum on the left side of the equality \reff{B04}
consider the generating function
\begin{equation}\label{B05}
\mathfrak{F}(x)=\sum_{n=0}^{\infty}\left[\sum_{m=0}^{\infty}(-1)^m C_{m+1}C_{n+m}^{2m+5}\right] x^n
\end{equation}
We rewrite $\mathfrak{F}(x)$ as follows
\begin{multline}\label{B06}
\mathfrak{F}(x)=\sum_{m=0}^{\infty}(-1)^m C_{m+1}\left[\sum_{n=0}^{\infty}C_{n+m}^{2m+5} x^n \right]=\\
\sum_{m=0}^{\infty}(-1)^m C_{m+1}x^{m+5}\left[\sum_{n=0}^{\infty}C_{n+m}^{2m+5} x^{n-m-5} \right].
\end{multline}

Taking into account  that for $m+5>n$ we have $C_{n+m}^{2m+5}=0$, we rewrite this equality as
\begin{equation}\label{B07}
\mathfrak{F}(x)=\sum_{m=0}^{\infty}(-1)^m C_{m+1}x^{m+5}\left[\sum_{n=m+5}^{\infty}C_{n+m}^{2m+5} x^{n-m-5} \right].
\end{equation}
From the power series
\begin{equation*}
\frac{1}{(1-x)^q}=\sum_{r=0}^{\infty} C_{q+r-1}^{q-1} x^r,
\end{equation*}
for $q=2m+6$ we obtain
\begin{equation}\label{B08}
\sum_{r=0}^{\infty} C_{2m+r+5}^{2m+5} x^{r}=\frac{1}{(1-x)^{2m+6}},
\end{equation}
or for $r=n-m-5$
\begin{equation}\label{B09}
\sum_{n=m+5}^{\infty} C_{n+m}^{2m+5} x^{n-m-5}=\frac{1}{(1-x)^{2m+6}}.
\end{equation}
Substituting \reff{B09} in \reff{B07}, we get
\begin{multline}\label{B10}
\mathfrak{F}(x)=\sum_{m=0}^{\infty}(-1)^m C_{m+1}x^{m+5}\frac{1}{(1-x)^{2m+6}}=\\
\frac{x^5}{(1-x)^6}\sum_{m=0}^{\infty}(-1)^m C_{m+1}\frac{x^m}{(1-x)^{2m}}=\\
\frac{x^5}{(1-x)^6}\frac{(1-x)^2}{-x}\sum_{m=0}^{\infty}C_{m+1}\left[\frac{(-x)}{(1-x)^{2}}\right]^{m+1}=\\
-\frac{x^4}{(1-x)^4}\sum_{m=0}^{\infty}C_{m+1}\left[\frac{(-x)}{(1-x)^{2}}\right]^{m+1}.
\end{multline}
Denoting $p=m+1$, we obtain
\begin{equation}\label{B11}
\mathfrak{F}(x)=\frac{-x^4}{(1-x)^4}\left[\sum_{p=1}^{\infty}C_{p}
\left(\frac{(-x)}{(1-x)^{2}}\right)^{p}+C_0-C_0\right],
\end{equation}
or, taking into account that $C_0=1$,
\begin{equation}\label{B12}
\mathfrak{F}(x)=\frac{-x^4}{(1-x)^4}\left[\sum_{p=0}^{\infty}C_{p}
\left(\frac{(-x)}{(1-x)^{2}}\right)^{p}\right]+\frac{x^4}{(1-x)^4}.
\end{equation}
Using the generating function for Catalan numbers
\begin{equation}\label{09}
\sum_{p=0}^\infty C_py^p=\frac{1-\sqrt{1-4y}}{2y},
\end{equation}
and the Maclaurin series expansion
of the function $\frac{1}{(1-x)^q}$ for $q=4$, we have
\begin{multline}\label{B13}
\mathfrak{F}(x)=\frac{-x^4}{(1-x)^4}(1-x)+\frac{x^4}{(1-x)^4}=\frac{x^4}{(1-x)^4}(1-1+x)=\frac{x^5}{(1-x)^4}=\\
x^5\left[\sum_{r=0}^{\infty}C_{r+3}^{3}x^r\right]=\sum_{r=0}^{\infty}C_{r+3}^{3}x^{r+5}=
\sum_{n=5}^{\infty}C_{n-2}^{\,n-5}x^{n}=\sum_{n=0}^{\infty}C_{n-2}^{\,n-5}x^{n},
\end{multline}
where $n=r+5$ and we take into account that $C_{n-2}^{\,n-5}=0$ for $n<5$. Thus \reff{B04},
and hence \reff{B01}, are proved.
\bigskip

{$\mathbf{2}$} Let's prove the second of the relations \reff{67}, namely
\begin{equation}{\label{D01}}
(-1)^{k-1}C_{k}^{k-1} + \sum_{t=0}^{k-2}(-1)^{k-2-t}C_t C_{k+1+t}^{k-1-t}=C_{k-1}.
\end{equation}

Note that for $t=k-1$ we have $(-1)^{k-2-t}=(-1)$ and $C_{k+t+1}^{k-t-1}=C_{2k}^{0}=1$. Thus we can
rewrite \reff{D01} as
\begin{equation}\label{D02}
\sum_{t=0}^{k-1}(-1)^{k-2-t} C_{t}C_{k+t+1}^{k-t-1}=(-1)^{k-2}C_{k}^{k-1}.
\end{equation}
Dividing by $(-1)^{k-2}$, and taking into account that
 $C_{k+t+1}^{k-t-1}=C_{k+t+1}^{2t+2}$, we obtain
\begin{equation}\label{D03}
\sum_{t=0}^{k-1}(-1)^{t} C_{t}C_{k+t+1}^{2t+2}=C_{k}^{k-1}.
\end{equation}
Because for $t\geq k$  we have $C_{k+t+1}^{2t+2}=0$, we can rewrite \reff{D03} in the form
\begin{equation}\label{D04}
\sum_{t=0}^{\infty}(-1)^{t} C_{t}C_{k+t+1}^{2t+2}=C_{k}^{1}.
\end{equation}
If we denote $n=k+1$ we obtain
\begin{equation}\label{D05}
\sum_{t=0}^{\infty}(-1)^{t} C_{t}C_{n+t}^{2t+2}=C_{n-1}^{1}.
\end{equation}
So we must prove this formula. To this end, we introduce a generating function
\begin{equation}\label{D05a}
\mathfrak{F}(x)=\sum_{n=0}^{\infty}\left[\sum_{t=0}^{\infty}(-1)^{t} C_{t}C_{n+t}^{2t+2}\right]x^n=
\sum_{t=0}^{\infty}(-1)^{t} C_{t}x^t\left[\sum_{n=0}^{\infty}C_{n+t}^{2t+2}x^{n-t}\right].
\end{equation}
Then we have
\begin{equation}\label{D06}
\mathfrak{F}(x)=\sum_{t=0}^{\infty}(-1)^{t} C_{t}x^{t+2}\left[\sum_{n=0}^{\infty}C_{n+t}^{2t+2}x^{n-t-2}\right].
\end{equation}
Since for $n<t+2$ we have $\sum_{n=0}^{\infty} \Rightarrow \sum_{n=t+2}^{\infty}$, we have
\begin{equation}\label{D07}
\mathfrak{F}(x)=\sum_{t=0}^{\infty}(-1)^{2+t} C_{t}x^{t+2}\left[\sum_{n=t+2}^{\infty}C_{n+t}^{2t+2}x^{n-t-2}\right].
\end{equation}
Then for $q=2t+3$ we have
\begin{equation}\label{D08}
\frac{1}{(1-x)^q}=\sum_{r=0}^{\infty}C_{q+r-1}^{q-1}x^r=\sum_{r=0}^{\infty}C_{2t+r+2}^{2t+2}x^r.
\end{equation}
Replacement $r=n-t-2$ gives
\begin{equation}\label{D09}
\frac{1}{(1-x)^q}=\sum_{n=t+2}^{\infty}C_{n+t}^{2t+2}x^{n-t-2}=\frac{1}{(1-x)^{2t+3}}.
\end{equation}
Then
\begin{equation}\label{D10}
\mathfrak{F}(x)=\sum_{t=0}^{\infty}(-1)^{t} C_{t}x^{t+2}\frac{1}{(1-x)^{2t+3}}=
\frac{x^2}{(1-x)^3}\sum_{t=0}^{\infty}C_{t}\left(\frac{-x}{(1-x)^{2}}\right)^t.
\end{equation}
Considering that (see \reff{09})
\begin{equation}\label{D11}
\sum_{t=0}^{\infty}C_{t}\left(\frac{-x}{(1-x)^{2}}\right)^t=1-x,
\end{equation}
we obtain
\begin{equation}\label{D12}
\mathfrak{F}(x)=\frac{x^2}{(1-x)^3}(1-x)=\frac{x^2}{(1-x)^2}=x^2\sum_{r=0}^{\infty}C_{r+1}^{1}x^r=
\sum_{r=0}^{\infty}C_{r+1}^{1}x^{r+2}.
\end{equation}
Replacing $n=r+2$ we get
\begin{equation}\label{D13}
\mathfrak{F}(x)=\sum_{n=2}^{\infty}C_{n-1}^{1}x^{n}=\sum_{n=0}^{\infty}C_{n-1}^{1}x^{n}.
\end{equation}
So
\begin{equation}\label{D14}
\sum_{t=0}^{\infty}(-1)^{t} C_{t}C_{n+t}^{2t+2}=C_{n-1}^{1},
\end{equation}
which was exactly what we needed to prove.

\section*{Appendix 2}
Let us prove \reff{48}:
\begin{equation}\label{D15}
\sum_{q=0}^{k-1}(-1)^{k-q-1}C_q C_{s+q}^{k-1-q}=(-1)^{k-1} C_{s-1}^{k-1},
\quad 1\leq k \leq s.
\end{equation}
or
\begin{equation}\label{D16}
\sum_{q=0}^{k-1}(-1)^{k-q}C_q C_{s+q}^{k-1-q}=(-1)^{k} C_{s-1}^{k-1}.
\end{equation}
Note that for $k-1<q$ we have $C_{s+q}^{k-1-q}=0$, we can rewrite \reff{D16} as
\begin{equation}\label{D17}
\sum_{q=0}^{\infty}(-1)^{k-q}C_q C_{s+q}^{k-1-q}=(-1)^{k} C_{s-1}^{k-1}.
\end{equation}
Consider generating function
\begin{equation}\label{D18}
\mathfrak{F}(x)=\sum_{s=0}^{\infty}x^s\left[\sum_{q=0}^{\infty}(-1)^{k-q}C_q C_{s+q}^{k-1-q}\right].
\end{equation}
We have
\begin{multline}\label{D18a}
\mathfrak{F}(x)=\sum_{q=0}^{\infty}(-1)^{k-q}C_q\left[\sum_{s=0}^{\infty}C_{s+q}^{k-1-q}x^s\right]=\\
\sum_{q=0}^{\infty}(-1)^{k-q}x^{k-2q-1}C_q\left[\sum_{s=0}^{\infty}C_{s+q}^{k-1-q}x^{s-k+2q+1}\right]
\end{multline}
Because for $k-1-q<s+q$ we have $C_{s+q}^{k-1-q}=0$ the last relation can be rewritten as
\begin{equation}\label{D19}
\mathfrak{F}(x)=\sum_{q=0}^{\infty}(-1)^{k-q}x^{k-2q-1}C_q\left[\sum_{s=k-1-2q}^{\infty}C_{s+q}^{k-1-q}x^{s-k+2q+1}\right]
\end{equation}
The sum in square brackets is the Maclaurin series expansion of the expression $(1-x)^{-p}$ for $p=k-q$:
\begin{equation}\label{D20}
\frac{1}{(1-x)^p}=\sum_{r=0}^{\infty}C_{p+r-1}^{p-1}\,x^r.
\end{equation}
So we obtain
\begin{equation}\label{D21}
\mathfrak{F}(x)\!=\!\sum_{q=0}^{\infty}(-1)^{k-q}x^{k-2q-1}C_q\frac{1}{(1-x)^{k-q}}\!=\!
(-1)^k\frac{x^{k-1}}{(1-x)^k}\sum_{q=0}^{\infty}C_q\left(-\frac{1-x}{x^2}\right)^q\!,
\end{equation}
or for $y=-\frac{1-x}{x^2}$
\begin{equation}\label{D22}
\mathfrak{F}(x)=(-1)^k\frac{x^{k-1}}{(1-x)^k}\sum_{q=0}^{\infty}C_qy^q=
(-1)^k\frac{x^{k-1}}{(1-x)^k}\frac{1-\sqrt{1-4y}}{2y},
\end{equation}
where we use the generating function for Catalan numbers \reff{09}. Because for $y=-\frac{1-x}{x^2}$
\begin{equation}\label{D23}
\frac{1-\sqrt{1-4y}}{2y}=x,
\end{equation}
we finally obtain
\begin{equation}\label{D24}
\mathfrak{F}(x)=(-1)^k\sum_{q=0}^{\infty}C_{k-r-1}^{x+s}.
\end{equation}
Thus the coefficient at $x^s$ is equal
\begin{equation}\label{D25}
(-1)^kC_{k+(s-k)-1}^{k-1}=(-1)^kC_{s-1}^{k-1},
\end{equation}
that's what we needed to prove.

\end{document}